\documentclass[aps,prl,reprint,superscriptaddress,showpacs]{revtex4-2}

\usepackage{graphicx}
\usepackage{epsfig} % use this package to include EPS format figures	
\usepackage{float}
\usepackage{amsmath}
\usepackage{amsfonts}%
\usepackage[utf8]{inputenc}
\usepackage[english]{babel} %francais, polish, spanish, ...
\usepackage[T1]{fontenc}
\usepackage{color}
\usepackage[pdftex,colorlinks]{hyperref}
\usepackage{MnSymbol}

\begin{document}

\title{First experimental observation of zonal flows in the optimized stellarator Wendelstein 7-X}
\author{D. Carralero}
\affiliation{CIEMAT, Avenida Complutense, 40, Madrid, Spain}
\author{T. Estrada}
\affiliation{CIEMAT, Avenida Complutense, 40, Madrid, Spain}
\author{J. M. García-Regaña}
\affiliation{CIEMAT, Avenida Complutense, 40, Madrid, Spain}
\author{E. Sánchez}
\affiliation{CIEMAT, Avenida Complutense, 40, Madrid, Spain}
\author{T. Windisch}
\affiliation{Max-Planck-Institut für Plasmaphysik, D-17491, Greifswald, Germany}
\author{A. Alonso}
\affiliation{CIEMAT, Avenida Complutense, 40, Madrid, Spain}
\author{E. Maragkoudakis}
\affiliation{CIEMAT, Avenida Complutense, 40, Madrid, Spain}
\author{C. Brandt}
\affiliation{Max-Planck-Institut für Plasmaphysik, D-17491, Greifswald, Germany}
\author{K. J. Brunner}
\affiliation{Max-Planck-Institut für Plasmaphysik, D-17491, Greifswald, Germany}
\author{C. Gallego-Castillo}
\affiliation{DAVE(ETSIAE), Universidad Politécnica de Madrid, Plaza Cardenal Cisneros 3, Madrid, Spain}
\author{K. Rahbarnia}
\affiliation{Max-Planck-Institut für Plasmaphysik, D-17491, Greifswald, Germany}
\author{H. Thienpondt}
\affiliation{CIEMAT, Avenida Complutense, 40, Madrid, Spain}
\author{the Wendelstein 7-X Team}
\noaffiliation
\date{\today}

\begin{abstract}
 
In this work, we present the first experimental evidence of the presence of zonal flow (ZF) structures in the optimized stellarator Wendelstein 7-X. Using an assortment of diagnostics, flux surface-uniform, electrostatic flow oscillations have been measured, showing a radial scale in the range of tens of ion gyroradii. Such measurements show remarkable agreement with the ZF predicted by local and global non-linear gyrokinetic simulations. These results represent the first direct measurement of ZF in a large stellarator, suitable for the validation of models in reactor relevant conditions.

\end{abstract}
%PACS:  52.35.Ra(Plasma turbulence) 52.55.Hc (stellarators), electrostatic waves and oscillations, 52.35.Fp), 47.27.-i (Turbulent flows)
\maketitle

\textit{Introduction -} Zonal Flows (ZF) are a universal characteristic of stratified turbulent systems which can be found in many different natural media such as oceans, planetary atmospheres or the solar convection zone \cite{Zonal_Jets}, where they manifest as system-sized flows perpendicular to the stratification. ZF can as well be found in some laboratory settings such as magnetic fusion experiments, in which high pressure plasmas are confined by means of strong magnetic fields arranged in nested flux surfaces. Plasma parameters are roughly constant across these surfaces, so ZF take the form of surface-uniform oscillations of the $\textbf{E}\times\textbf{B}$ flow, $u_{\textrm{E}}$, perpendicular both to the magnetic field, $\textbf{B}$, and the electric field, $\textbf{E}$, normal to the surface and thus aligned with the stratification direction \cite{Diamond05}. In this context, ZF are excited through non-linear coupling of unstable micro-scale modes (typically drift wave turbulence, possibly amplified by other mechanisms such as interaction with fast particles \cite{diSiena21}) and are expected to play an important role in confinement: While not directly causing advection perpendicular to flux surfaces, they are supposed to regulate cross-field turbulent transport by a number of mechanisms including suppression of turbulence by shear flows \cite{Lin98}, stabilization of drift modes by inverse cascade energy transfer  \cite{Diamond05}, or govern saturation by the activation of large-scale damped modes \cite{Terry15}. This central role in turbulence regulation makes a proper characterization of ZF in reactor-relevant plasmas of paramount importance for the development of a fusion reactor, and particularly so for a stellarator reactor: Initial operation of Wendelstein 7-X (W7-X) \cite{Wolf17} has demonstrated that optimization can reduce collisional transport associated to the stellarator-characteristic three dimensionality of the magnetic field (usually termed ``neoclassical transport'') to acceptable levels, while retaining all their advantages over tokamaks, such as the absence of disruptions, easier access to steady-state operation, etc.  \cite{Beidler19}. However, the first experimental campaigns have shown that turbulence typically dominates transport and only in a handful of scenarios in which it can be suppressed, confinement levels consistent with reactor requirements can be achieved \cite{Dinklage18,Bozhenkov19}. This has triggered a widespread effort to characterize and model turbulent transport in stellarators, notably including ZF, which have been proposed as the cause of the enhanced turbulence present in W7-X compared to the similarly sized LHD heliotron \cite{Warmer21}. So far, this effort was greatly hampered by the fact that no ZF had ever been detected in any large stellarator. In this work, we present the first experimental detection and  characterization of ZFs in W7-X, thereby setting the foundation for a new line of research on this key mechanism. Since this device is expected to achieve fusion relevant plasma parameters when fully powered \cite{Wolf17}, such new line might eventually be critical for the validation of our current models for the regulation of transport in a future stellarator reactor.

\begin{figure}[h!]
	\centering
	\includegraphics[width=\columnwidth]{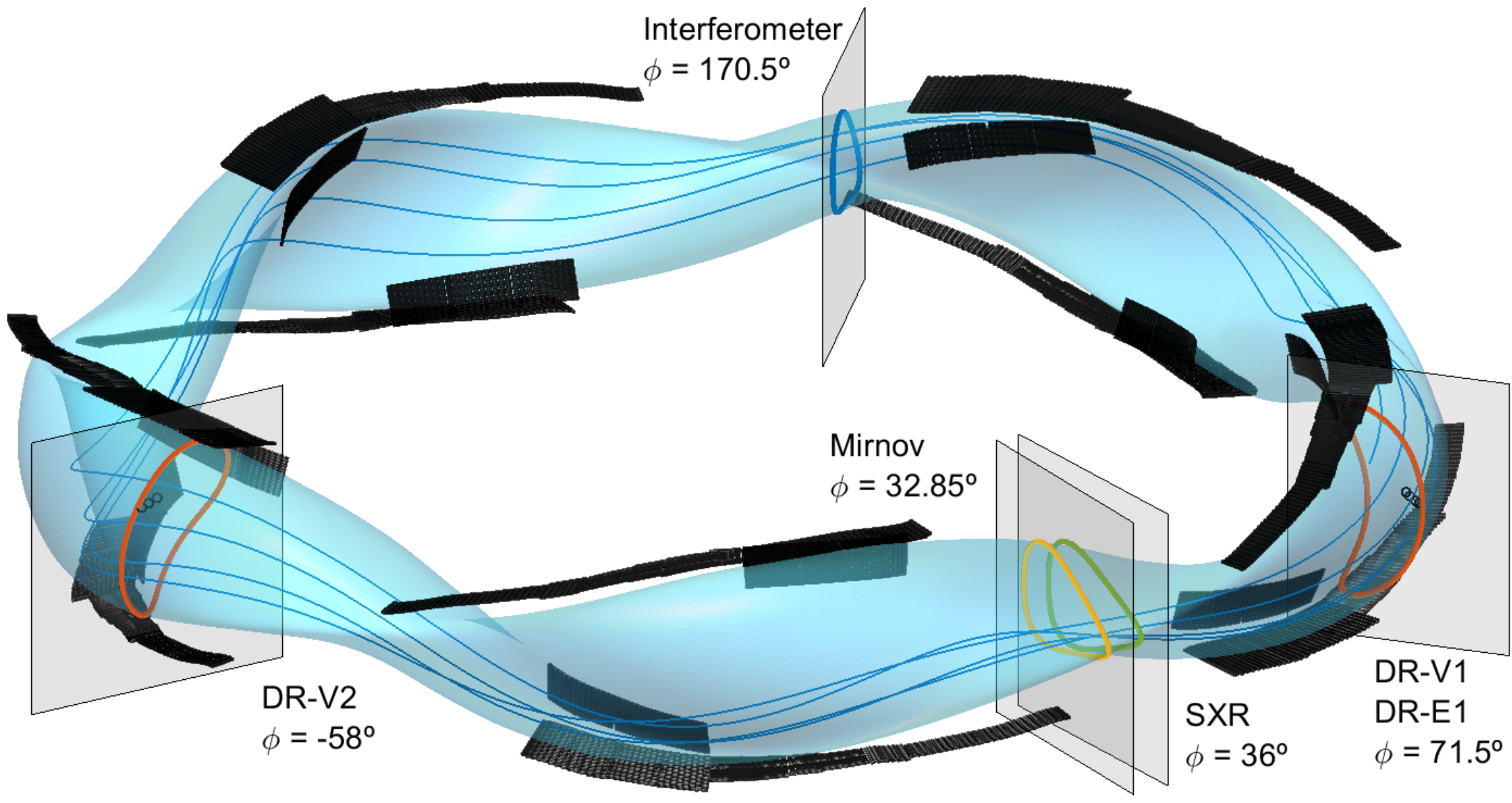}
	\caption{Experimental layout at W7-X. Last closed flux surface and divertors are indicated as blue shade and black structures. The cross section at each diagnostic position is indicated as a colored curve. DR measurement areas are highlighted by black circles. Blue line shows two toroidal turns in each direction  of a flux tube crossing the $\rho = 0.7$ measuring position of DR-V1.}
	\label{fig:1}
\end{figure}

\begin{figure}
	\centering
	\includegraphics[width=\columnwidth]{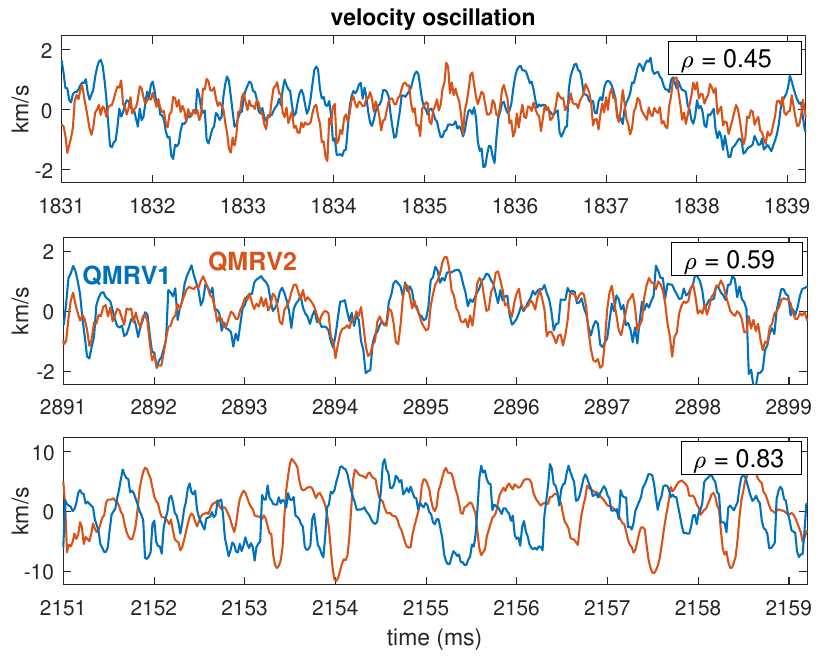}
		\includegraphics[width=\columnwidth]{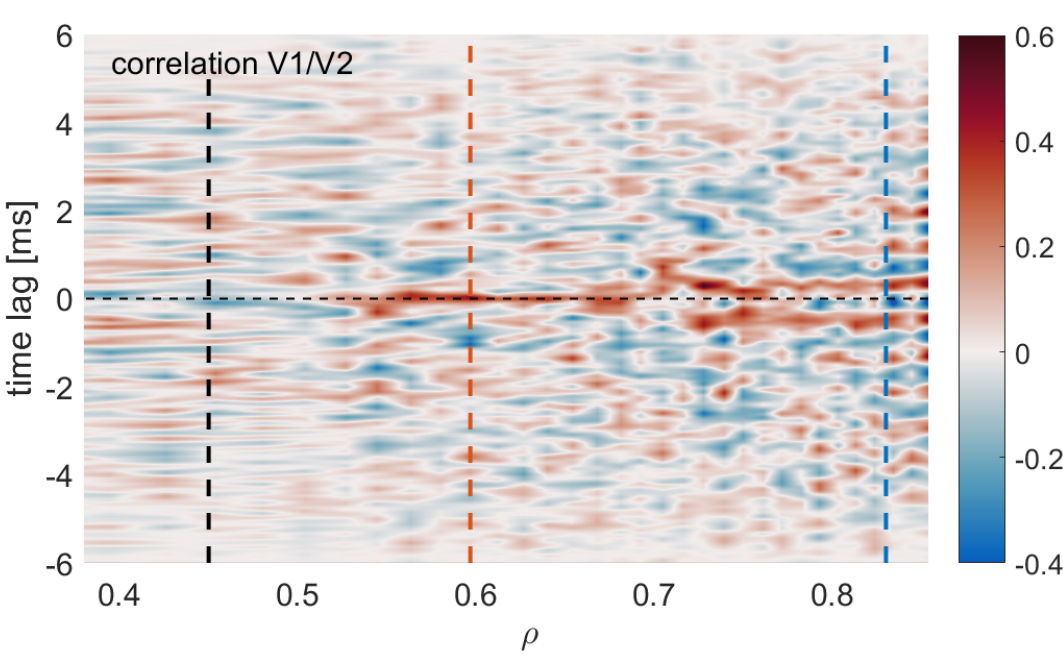}	
		
	\caption{Dual reflectometer signal analysis from reference discharge $20230215.051$. Top, $u_\perp$ signals from DR-V1 and DR-V2 at three radial positions. DC component has been filtered out in them. Bottom, cross-correlation of the two signals. Each position at the x-axis corresponds to a common DR beam frequency, resulting in a different radial location for the measurement (as indicated by the black circles in Figure \ref{fig:1}.) Positions from the top plot are indicated by a dashed line.}
	\label{fig:2}
\end{figure}

\textit{Experiments -} Empirical observation of ZF is a daunting endeavor: Given the flux surface uniformity of ZF, their most distinctive experimental signature is the presence of long range correlations (LRC): significant cross-correlation levels between measurements of velocity oscillations carried out in distant, magnetically-disconnected positions of the same flux surface. This is a demanding technique, as it typically requires duplicated fluctuation diagnostics such as Langmuir probes or HIBP, which has been used for the detection of ZF in most studies in the literature \cite{Fujisawa04,Pedrosa08,Alonso17}. In W7-X, a dual Doppler reflectometer (DR) system has been developed and recently commissioned with the main objective of carrying out this kind of measurements: as can be seen in Figure \ref{fig:1}, each DR (named DR-V1 and DR-V2) is installed in one of two sections located $129^{\circ}$ toroidally apart, and their measurement regions are not directly connected by a field line. These diagnostics measure the Doppler-displaced back-scattered component of a V-band microwave beam from which the local perpendicular velocity, $u_\perp \simeq u_{\textrm{E}}$, can be calculated. More details on the DR system and the validity of this approximation can be found in \cite{Carralero20,Estrada21,Windisch22,Windisch23}. In this work, we analyze stationary plasmas of intermediate density ($\bar{n}_{\textrm{e}} \simeq 6.5 \ 10^{19}$ m$^{-2}$) on which $5$ MW of electron-cyclotron resonance heating is applied. Both DR carry out synchronized radial sweeps with frequency being changed in discrete 1 GHz steps of 10 ms duration to probe a range of radial positions. This  frequency synchronization results also in a radial position match up, as confirmed by ray tracing analysis carried out with TRAVIS \cite{Marushchenko14}. The 10 MHz sampled complex amplitude signals are binned and Fourier transformed in order to obtain a 50 kHz measure of the Doppler peak frequency oscillation, from which a temporal trace of $u_\perp$ can be calculated for each DR \cite{Estrada12}. Finally, the cross-correlation of the two $u_\perp$ signals is calculated in order to obtain the LRC. A characteristic result of this analysis is shown in the top plot of Figure \ref{fig:2}, which shows representative examples of the $u_\perp$ signals at three different radial regions (indicated by $\rho=r/a$, the effective minor radius coordinate normalized by the minor radius of the plasma): For $\rho \simeq 0.45$ (``inner core'') both signals are clearly different, although this is most likely due to the quality of the DR-V2 signal being strongly reduced for $\rho < 0.5$. Instead, a high correlation can be appreciated even by the naked eye at $\rho \simeq 0.6$ (``outer core''), similar to that of the HIBP measurements reported in \cite{Fujisawa04} (although it should be noted that no band-filter has been used here to improve the match). Finally, signals in $\rho \simeq 0.85$ (``edge'') display substantially stronger oscillations with peaks of up to $10$ km/s, which seem to be more regular than the ones found at the core and have roughly opposing phases in each channel. These three regions can be better appreciated in the bottom plot of Figure \ref{fig:2}, in which the LRC over the full V-band is shown, covering a range of measurements for $\rho = [0.4-0.85]$. In it, the edge region appears for $\rho > 0.7$, where a strongly coherent mode is marked by the presence of regular positive and negative lobes. This mode has a well defined frequency of $1.2$ kHz, and displays a clear anti-phase between DR-V1 and V2. At the outer core ($0.55 < \rho < 0.7$), the lack of lobes reveals a global correlation caused by some non-coherent oscillation, which ranges between a few hundred Hz and $2$ kHz. Interestingly, a similar modulation can be seen in estimated flows from line-integrated PCI measurements \cite{Baener22}. In this region, the maximum value at time lag $\tau= 0$ indicates a zero-phase between the two DR channels. Finally, while LRC are not detected in the inner core, no physical interpretation from this fact is proposed due to the aforementioned low DR-V2 signal level for $\rho < 0.5$. 

\begin{figure}[h!]
%	\centering
	\includegraphics[width=\columnwidth]{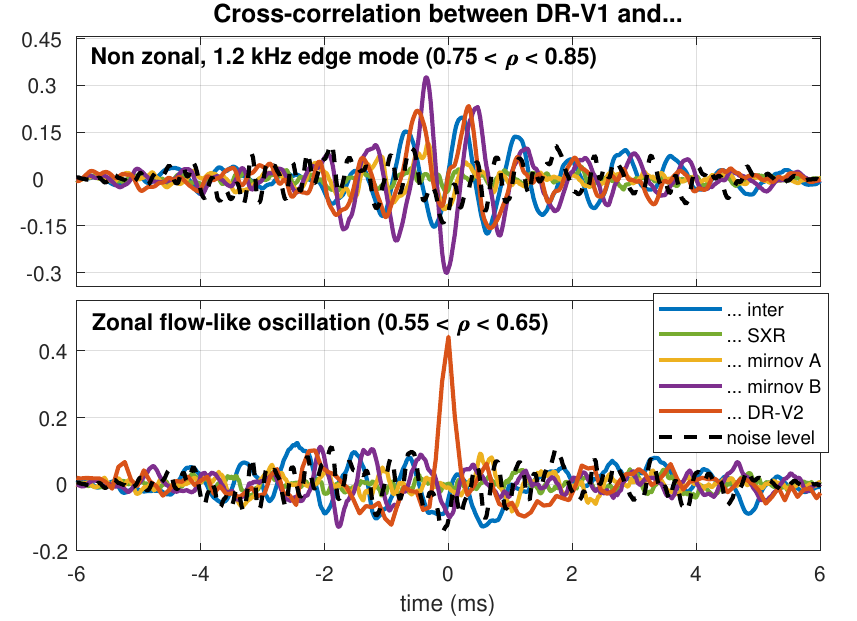}	
	\caption{Top/bottom plots show the cross-correlation of DR-V1 measurements at the edge/core regions with other diagnostics. Noise level is represented as a dashed line.}
	\label{fig:3}
	
	\centering
	\includegraphics[width=\columnwidth]{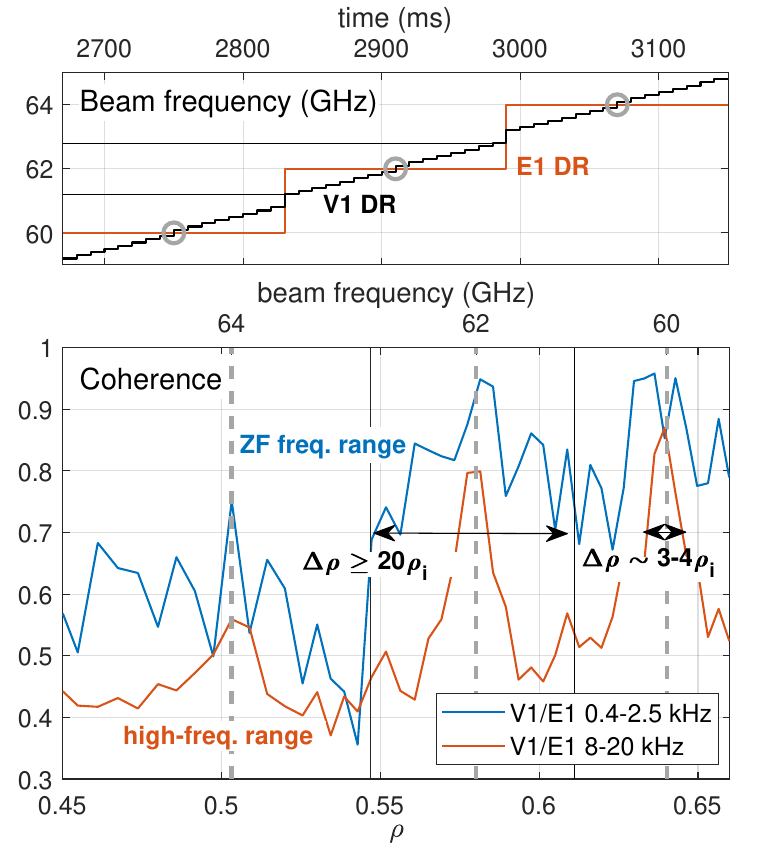}	
	\caption{Radial correlation study. Top: frequency sweeps of all three DR systems for a typical 500 ms ramp. Bottom: coherence between DR-V1 and DR-E1 signals at different radii and frequency ranges. Vertical dashed lines indicate the fixed E1 frequencies/radii for which both DR are measuring at the same position. As a guide to the eye, the limits of the V1 scan around $62$ GHz are highlighted in both figures as solid black lines.}
	\label{fig:4}
\end{figure}

The DR signature of the mode observed at the edge does not correspond to the one that could be expected for a ZF, as the two remote channels exhibit a clear phase delay. Instead, its frequency and radial location are very reminiscent of an electromagnetic quasi-coherent edge mode, well documented in the literature with a range of diagnostics (see, eg. \cite{Ballinger21}, where the same 1.2 kHz mode appears in a very similar plasma scenario). In order to confirm the identification of the edge mode detected in this work, the velocity oscillations measured by the DR are cross-correlated with measurements from several of those diagnostics, placed in different toroidal locations, as indicated in Figure \ref{fig:1}: First, a fast line integrated density signal from the interferometer \cite{Brunner18}; second, soft X-ray (SXR) measurements from a range of cords corresponding to the radial region covered by DR \cite{Brandt17}; finally, magnetic field fluctuations measured by a number of Mirnov coils placed at different poloidal positions \cite{Rahbarnia18}. The results can be seen in Figure \ref{fig:3}, where cross-correlations of each of these diagnostics and DR-V1 oscillations from the edge and outer core regions are presented. For clarity, only one SXR chord (2A-20, crossing the whole plasma section) and two Mirnov signals (11-120 and 11-440,  separated poloidally roughly 90$^\circ$) have been selected as the most representative. Also, the average cross-correlation between DR-V1 and DR-V2 in the $0.55 < \rho < 0.65$ and $0.75 < \rho < 0.85$ region is represented in black. Finally, the cross-correlation between the two DR in the inner core (where DR-V2 is pure noise) is represented in both plots as a reference of the background noise level of the analysis. As can be seen in the figure, this mode is clearly beyond the noise level and present in all the surveyed quantities. As well, substantial phase differences can be seen between the various channels, including the two poloidally separated Mirnov coils, thus revealing non-uniform poloidal/toroidal structure. These results are consistent with previous reports and ratify that the edge correlation corresponds to the edge mode previously described in the literature. Also, they confirm the non-ZF nature of the edge mode: on top of the non-uniformity, no LRCs in density or current equivalent to those in flow are expected for a ZF \cite{Diamond05,Pedrosa08}.

\begin{figure}[t!]
	\centering
	\includegraphics[width=\columnwidth]{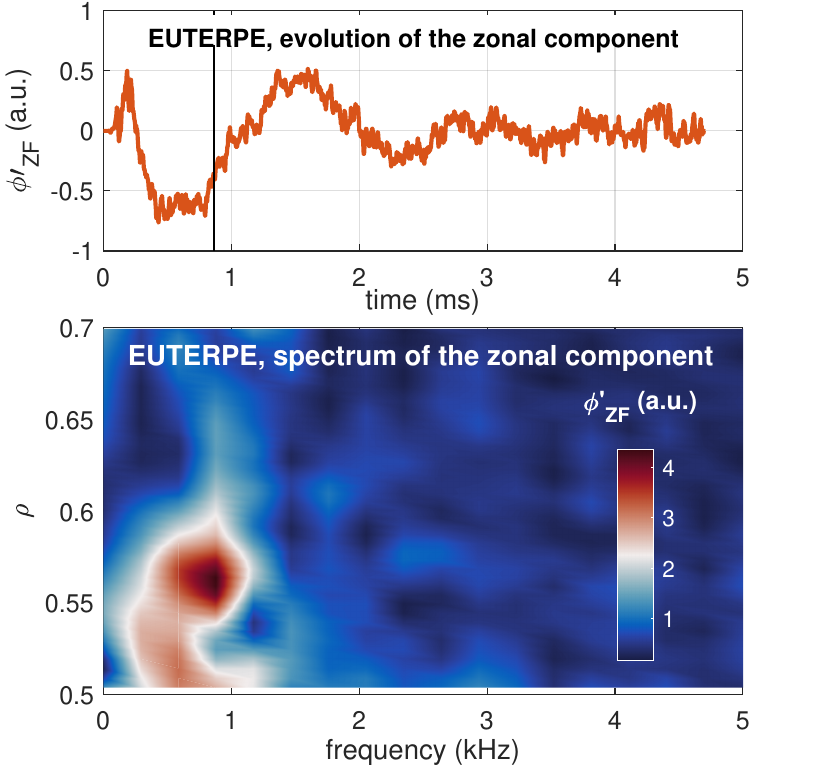}
	\caption{EUTERPE GK simulations. Top: Evolution of the zonal component of a typical run. Bottom: Power spectrum  vs radius and frequency for the zonal component.}
	\label{fig:5}
	\centering
	\includegraphics[width=.95\columnwidth]{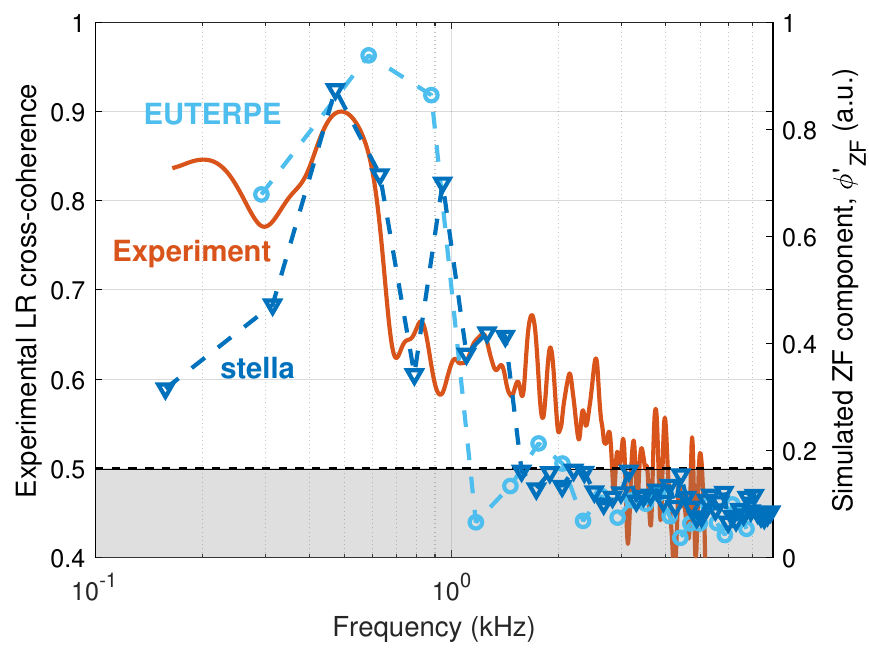}
	\caption{Comparison between GK simulations and experimental results. Grey shaded area indicates non-meaningful experimental cross-coherence.}
	\label{fig:6}
\end{figure}

On the contrary, when this analysis is carried out with DR signals from the  $0.55 < \rho < 0.65$ region, results are clearly consistent with a ZF signature: As can be seen in the lower plot of Figure \ref{fig:3}, the only cross-correlation above the noise level is the LRC between DR-V1 and V2, indicating that the non-coherent velocity oscillation is not only in phase at the two remote DR systems, but also not associated to the current or density fluctuations detected by the other diagnostics, as expected for the flux-surface-uniform, electrostatic oscillations of a ZF. Further indication of this can be found by a radial correlation analysis: To this end, a fine radial scan was carried out in a number of experiments using small frequency steps of 100 MHz in the synchronous frequency ramp of DR-V1 and V2. At the same time, a third DR system (DR-E1) measured at constant frequencies values at intermediate points of the scan (60, 62 and 64 GHz. See Figure \ref{fig:4} for a sketch of the frequency setup). DR-E1 is installed in the same port and antenna as DR-V1, meaning that both systems provide $u_\perp$ measurements separated by a radial distance related to their frequency difference (and from the exact same point when both frequencies coincide). Using this layout, the radial correlation length of velocity fluctuations can be obtained at the measurement location \cite{Schirmer07}. The results of such analysis are shown in the bottom plot of Figure \ref{fig:4}: first, the coherence spectrum between both signals is calculated for each combination of V1/E1 frequencies. Then, characteristic values are averaged in two frequency ranges: a low band, characteristic of the ZF oscillations ([0.4, 2.5] kHz, see Figure \ref{fig:6}), and a high band for which no ZF activity is expected according to the LRC spectrogram ([8-20 kHz]). Finally, the two averaged coherences are represented as a function of the DR-V1 measurement radial position. As could be expected, coherence is maximum for both curves at points where the radial scan crosses the long position steps of DR-E1 (indicated in the figure as three gray circles/dashed lines). However, the low frequency, ZF-relevant curve reaches higher coherence values and shows a much slower radial decay, keeping substantial coherence values of $\gamma \simeq 0.7$ at the edges of the frequency scan (marked by thin black lines and spanning $1.6$ GHz, corresponding to $\Delta \rho \simeq 0.05$). This can be interpreted as a lower bound for the radial correlation length which, taking minor radius $a=0.5$ m and $T_i \simeq 1\,\textrm{keV}$, would yield a minimum radial width of the ZF of $2.5$ cm or roughly $20$ times the ion gyroradius ($\rho_i$). This is in good agreement with the ZF expected radial correlation length of several tens of $\rho_i$ \cite{Diamond05}. By contrast, the radial correlation length of the high frequency band corresponds to a few $\rho_i$, more in line with the typical values for common drift-wave turbulence.

%\begin{figure}

%\end{figure}

\textit{Simulations -} In order to substantiate this experimental identification, measurements are compared to non-linear gyrokinetic (GK) simulations  carried out with realistic plasma profiles. First, radially global simulations with adiabatic electrons were carried out with EUTERPE \cite{Kleiber24} covering the $0.2 <\rho < 0.9$ domain. A period (one-fifth) of the full stellarator is simulated with a spatial resolution similar to that used in \cite{Sanchez23}, which allowed for a good convergence. Source term and weight smoothing are used as in \cite{Sanchez20} to keep the density and temperature profiles stable and reduce numerical noise in the (very long) nonlinear saturated state. The evolution of the zonal potential (defined as the flux surface-uniform component of the potential, $\phi_{\textrm{ZF}}$) is monitored in the region $0.5<\rho<0.7$, keeping aside two buffer regions to limit the effect of the inner and outer boundaries. From this, the radial derivative giving rise to the zonal flows, $\phi_{\textrm{ZF}}'$, is calculated. Second, simulations with kinetic electrons were performed with the code stella \cite{Barnes19} in flux tube geometry. In order to best approximate experimental observations, the radial position $\rho=0.6$ and the flux tube centered with respect to the bean-shaped poloidal cross section of the plasma were considered, extending the flux tube two turns toroidally (following a field line equivalent to the one depicted in Figure \ref{fig:1}). As before, the code calculates the $\phi_{\textrm{ZF}}'$ from which $u_{\textrm{E}}$ can be obtained (both $\phi_{\textrm{ZF}}'$ definitions are equivalent, although stella-specific details can be found in \cite{Gonzalez-Jerez24}). It must be noted that all these simulations are extremely intensive in computation resources, as exceptionally long times are required to resolve the sub-kHz ZF oscillations: Stella and EUTERPE consumed in the order of $2 \cdot 10^5$ and $10^6$ CPU-hours respectively for simulating around 5 ms (equivalent to roughly $5000 \times a/v_{\textrm{th,i}}$, with $v_{\textrm{th,i}}$ the thermal speed of the ions). As an example, the evolution of $\phi_{\textrm{ZF}}'$ at $\rho = 0.55$  obtained by EUTERPE for reference discharge $20230215.051$ is displayed on the top plot of Figure \ref{fig:5}. The bottom plot shows the spectrogram of  $\phi_{\textrm{ZF}}'$ over the $0.5 <\rho < 0.7$ radial range. As can be seen, the simulation correctly captures the radial localization of the ZF in the experiment (the $\rho < 0.6$ localization surrounded by an outer, activity-free region matches qualitatively with the zero phase LRC at Figure \ref{fig:2}) and indicates a peak frequency for the ZF in the sub-kHz range, which can be identified in the top plot of Figure \ref{fig:5} as well as in equivalent evolution plots obtained by stella. Finally, a direct comparison between both simulations and experiments is presented in Figure \ref{fig:6}: In it, the spectrum of the long range coherence between DR-V1 and V2 is calculated reducing the Nyquist frequency to some $5$ kHz in order to obtain several subsamples to average, thus ensuring that the minimum statistically meaningful coherence is $\gamma_{V_1,V_2} = 0.5$ \cite{Bendat} (the non-meaningful region below that value is indicated as a shaded area in the figure). As can be seen, the coherence peaks at around 500 Hz and remains significant up to around 3 kHz. This result is compared to the ZF component spectra from both simulations, taking $\rho = 0.55$ position in the case of EUTERPE as representative of the ZF activity displayed in Figure \ref{fig:5}. The three curves show a remarkable agreement in the structure and frequency of the peak, which appears respectively for $470$ and $580$ Hz for stella and EUTERPE. As well, qualitative agreement is reached on the spectral range for which the LRC are detected at the experiment (up to $3$ kHz) and simulations (up to ca. 2 kHz). 

\textit{Conclusions -} In conclusion, positive evidence for the existence of ZF in Wendelstein 7-X is presented for the first time: First, LRC analysis between two remote DR proves the existence of a global structure in $u_\perp$ oscillations measured in the $0.55 < \rho < 0.65$ region (deeper radii can not be probed due the loss of DR signal) and in a spectral range between a few hundred Hz and around $3$ kHz. This result is so clear that can be seen directly in the unfiltered raw data shown in Figure \ref{fig:2}. Second, comparison to other diagnostics reveals the electrostatic nature of the structure. This result clearly departs from the correlation to current and density fluctuations measured at the edge mode in $\rho \ge 0.75$, in good agreement with previous reports. Third, an additional analysis of the mode structure carried out with a third DR yields a radial size of at least $20$ ion gyroradii, much higher than that of conventional drift wave turbulence and consistent with theoretical expectations. Finally, the frequency spectra of the LRC shows a remarkable agreement with the predicted ZF component of independent non-linear gyrokinetic simulations, which were extraordinarily CPU-intensive in order to resolve the low frequencies involved. The groundbreaking importance of this result comes not just from the fact that theoretical predictions regarding a key feature of the turbulent system regulating transport in an optimized stellarator have been experimentally confirmed, but also from the first time detection of ZF in a machine capable of reactor-relevant plasmas, which opens a new line of turbulence experimental research and theory validation with direct implications for next generation devices.

\textit{Acknowledgements -} The authors acknowledge the entire W7-X team for their support. Simulations were carried out in Marconi and Marenostrum-IV supercomputers. This work has been funded by the Spanish Ministry of Science, Innovation and Universities under grant PID2021-125607NB-I00 and PID2021-123175NB-I00 and was carried out within the framework of the EUROfusion Consortium, funded by the European Union via the Euratom Research and Training Programme (Grant Agreement No 101052200 - EUROfusion). Views and opinions expressed are however those of the authors only and do not necessarily reflect those of the European Union or the European Commission. Neither the European Union nor the European Commission can be held responsible for them.

\end{document}